\documentclass{kapproc} 
\let\footnote\savefootnote
\let\footnotetext\savefootnotetext 
 
\kluwerbib

\RequirePackage{graphicx}%
\RequirePackage{epsf}%

\begin{document}

\articletitle{Where has all the neutral hydrogen gone?}

\author{Disney, M. J.}

\affil{Department of Physics and Astronomy, Cardiff University, Cardiff, CF24 3YB, UK}
\email{m.disney@astro.cf.ac.uk}

\author{Minchin, R. F.}

\affil{Department of Physics and Astronomy, Cardiff University, Cardiff, CF24 3YB, UK}
\email{r.minchin@astro.cf.ac.uk}


\begin{abstract}
Our extremely deep survey for extragalactic H{\sc i} (HIDEEP) finds 
no intergalactic gas clouds, and no galaxies with H{\sc i} at inferred 
global column-densities below $10^{20}$~cm$^{-2}$ when we could have 
detected such objects down to a completeness limit of $4\times
10^{18}$~cm$^{-2}$.  We speculate that low surface-density hydrogen is either
ionised or locked up in ``frozen discs'', i.e. structures where 
the local Ly-$\alpha$ is insufficient to excite the 21-cm transition to a 
spin-temperature above the cosmic background.  Such ``frozen discs'' might
be responsible for many QSOALSs.
\end{abstract}


\section{Introduction}
The very strong selection effects against finding low surface-brightness galaxies (LSBGs) and dwarfs in the optical make it imperative to carry out blind searches at 21-cm where redshift can discriminate between extragalactic and foreground hydrogen.  However, low surface-density and low surface-brightness systems might be expected to have low H{\sc i} column-densities ($N_{HI}$) which can only be reached by long integrations (Disney \& Banks 1997) -- irrespective of telescope size.  For instance, to reach $N_{HI} = 10^{18.5}$~cm$^{-2}$ when the gas is spread over 200 km\,s$^{-1}$, even for a source which fills the beam, requires integrations of $\sim 10^4$~s or more.  Such deep searches have never been possible before but we have taken advantage of the multibeam system at Parkes to integrate for 9000~s per point over a 32 square degree patch of sky, for 24 square degrees of which we have equivalently deep optical observations (reaching 26.5~R\,mags\,arcsec$^{-2}$). This HIDEEP survey (Minchin 2001, Minchin et al. 2002) fails to turn up any intergalactic clouds, or indeed any gas with a mean inferred column density less than $10^{20}$~cm$^{-2}$, when the survey was capable of reaching down to $4\times 10^{18}$~cm$^{-2}$ for galaxies with $\Delta V = 200$~km\,s$^{-1}$.  Why?

\section{The inferred column-density, $N_{HI}^\circ$}

Since we have not imaged most of our sources in H{\sc i} we can only calculate, following earlier authors, the inferred column-density, $N_{HI}^\circ$ by combining the measured H{\sc i} mass, $M_{HI}$, with an estimate of the hydrogen radius, $r_{HI}$, based on the measured optical radius $r_{eff}(R)$, where we have assumed
\begin{equation}
N_{HI}^\circ = \frac{M_{HI}}{m_H \pi r^2_{HI}}
\end{equation}
\noindent with $r_{HI} = 5 r_{eff}(R)$, taken from the earlier work of authors such as Salpeter \& Hoffman (1996).

\begin{figure}
\centering
\leavevmode
 \columnwidth=.45\columnwidth
 \includegraphics[width={\columnwidth}]{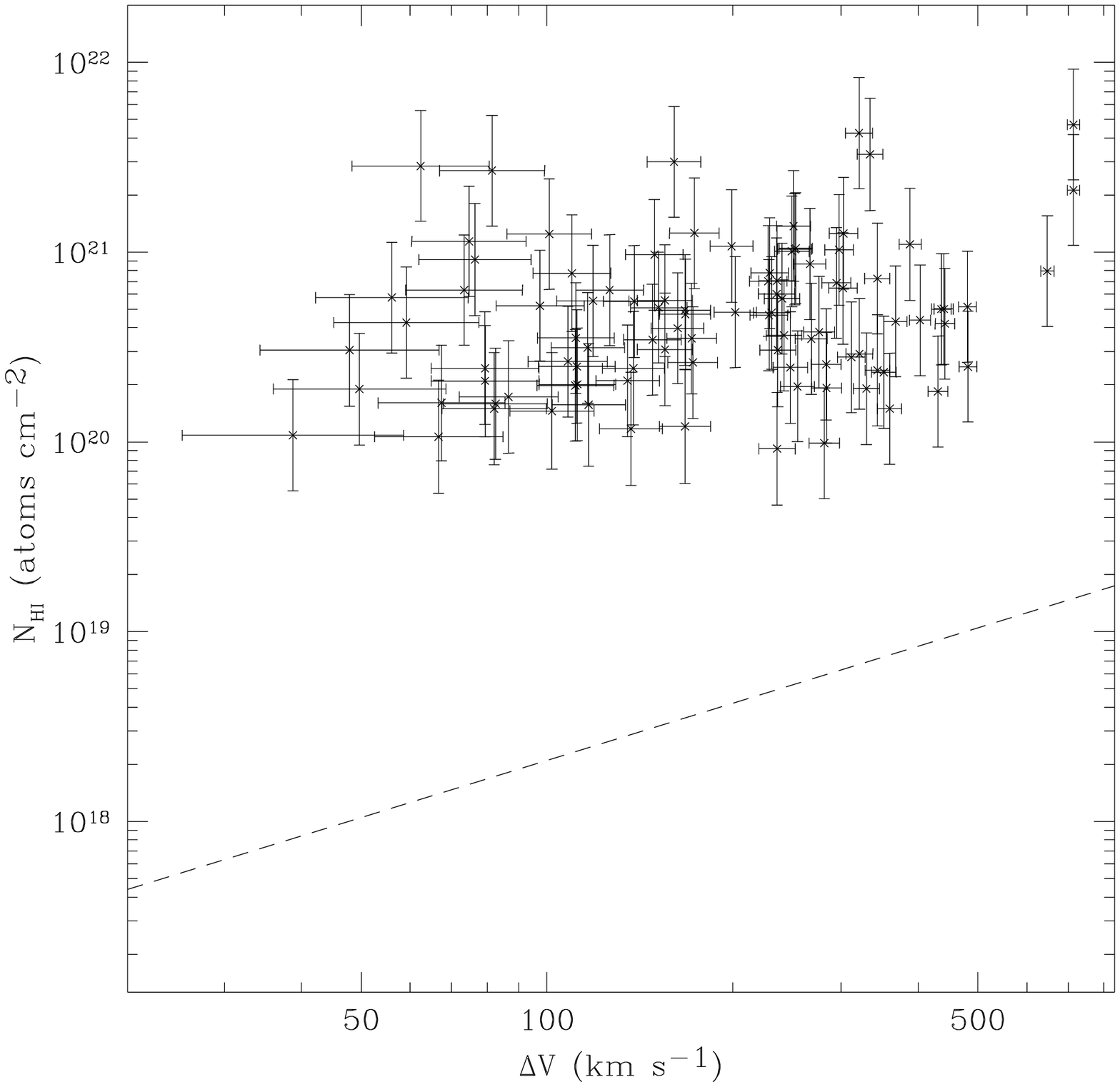}
 \hfil
 \includegraphics[width={\columnwidth}]{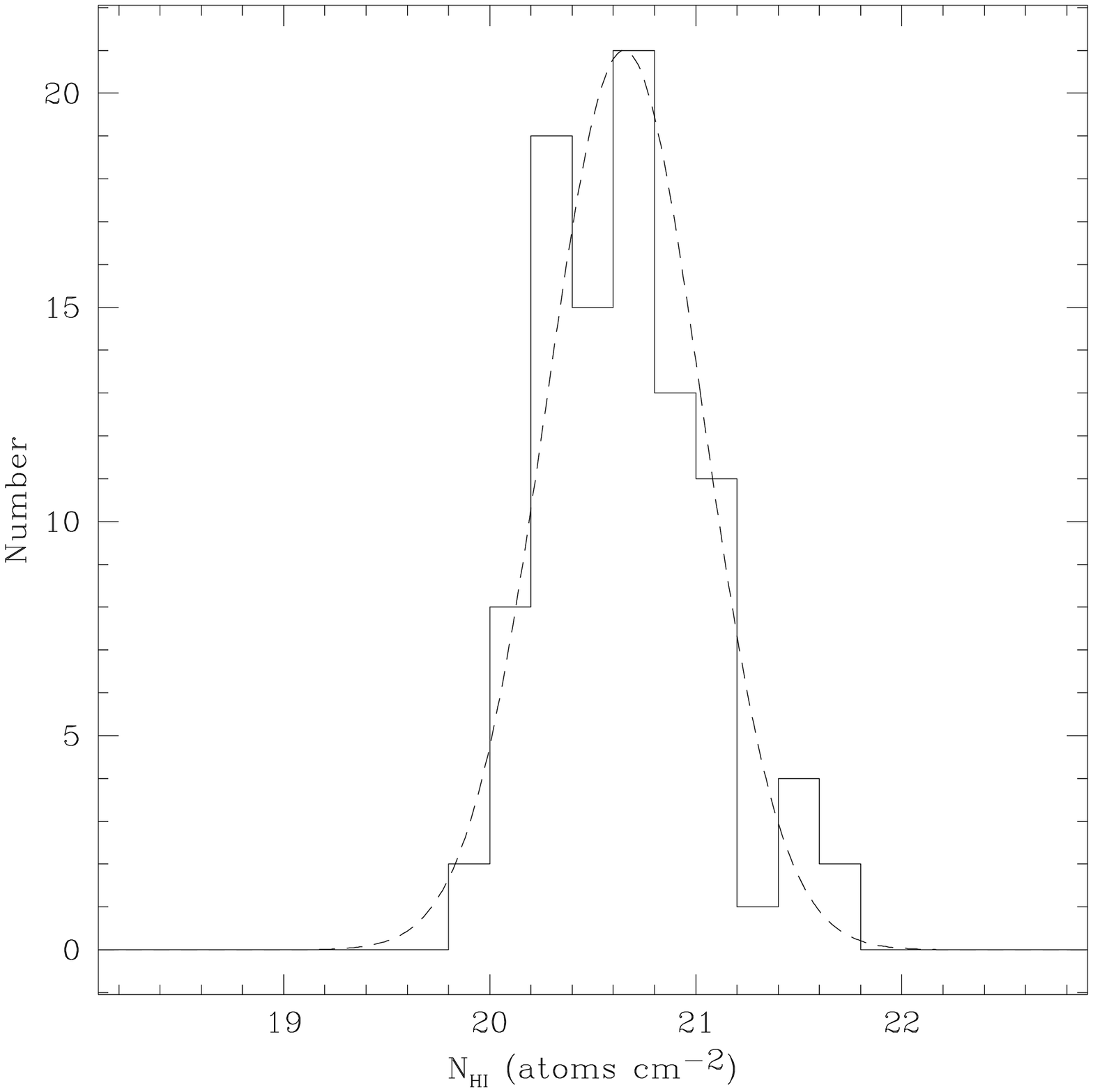}
\caption{(a) Inferred column-density as a function of velocity-width, dashed line shows detection limit (b) distribution of inferred column-densities.}
\end{figure}

Figure 1 shows the results for the 96 sources with uniquely-identified optical counterparts which lie in the overlap area between the H{\sc i} and optical surveys. Figure 1b in particular is dramatic.  Allowing for observational errors it says, in essence, that all galaxies have exactly the same inferred H{\sc i} column-density, i.e. $\log N_{HI}^\circ = 20.6\pm 0.3$.  A similar kind of result was reported by Giovanelli \& Haynes (1984) for optically selected galaxies.

This result could not be a selection effect because: (a) this is a blind survey (i.e. no selection of targets) and (b) our sensitivity to low column-density gas can be worked out retrospectively using the flux of the  weakest sources actually detected from
\begin{equation}
N_{HI}^{min} = 4.5\times 10^{20}\left(\frac{F_{HI}}{\Delta V \theta^2}\right)
\end{equation}
\noindent where $F_{HI}$ is the integrated H{\sc i} flux in Jy\,km\,s$^{-1}$, $\Delta V$ is the velocity width in km\,s$^{-1}$ and  $\theta$ is the beam FWHM in arcminutes.  This gives a limit for HIDEEP of $\sim 10^{18}$~cm$^{-2}$ at a velocity width of 100~km\,s$^{-1}$ showing that we should have easily detected low $N_{HI}$ systems.

\section{Possible astrophysical effects}

If one integrates the equation of radiative transfer for 21-cm radiation passing thorugh a uniform slab, then (e.g. Kulkarni \& Heiles 1988)
\begin{equation}
\Delta T \equiv T_B(\nu) - T_{bg} = \left(T_{spin} - T_{bg}\right)\left(1-e^{-\tau(\nu)}\right)
\end{equation}
\noindent where $T_B$ is a measure of the surface-brightness of the object in the line, $T_{spin}$ is the excitation temperature, and $\tau$ is the optical depth at frequency $\nu$.  If $\Delta T = 0$ the object is indistinguishable from the background.  It is straightforward to calculate that 
\begin{equation}
\tau(\nu) = \frac{5.5\times 10^{-18}}{T_{spin}(K)} \frac{N_{HI}({\rm cm}^{-2})}{\Delta V ({\rm km\,s}^{-1})}
\end{equation}
\noindent where $\Delta V$ is the velocity dispersion along the line of sight.  Clouds with $\Delta V \simeq 10$~km\,s$^{-1}$ have significant optical depth for $N_{HI}\geq 10^{21}$~cm$^{-2}$.

It is usually assumed (Field 1959, Kulkarni \& Heiles 1988) that, even where collisions are ineffective, the excitation (spin) temperature of H{\sc i} gas is maintained at the kinetic temperature by Ly-$\alpha$ photons.  This is so because neutral gas is so optically thick in the Ly-$\alpha$ line.  However at low enough densities this mechanism can fail and the upper level will then be populated only by interaction with the Cosmic radio background.  In the optically thin limit Equation (4) can be substituted into (3) when
\begin{equation}
\Delta T = \left(\frac{T_{spin} - T_{bg}}{T_{spin}}\right) 5.5\times 10^{-18}\frac{N_{HI}({\rm cm}^{-2})}{\Delta V ({\rm km\,s}^{-1})}
\end{equation}
\noindent so when $T_{spin}\rightarrow T_{bg}$, $\Delta T\rightarrow 0$ and such a slab or disc would become indistinguishable from the CBR.  Watson and Deguchi (1984) have calculated numerically the escape of Ly-$\alpha$ photons generated in the gas by the X-ray background.  They find that if
\begin{equation}
\frac{N_{HI} ({\rm cm}^{-2})}{\Delta V ({\rm km\,s}^{-1})} \leq 6\times 10^{19}
\end{equation}
\noindent then $\left(\frac{T_{spin} - T_{bg}}{T_{spin}}\right)$ in Equation (5) rapidly drops below 0.5 and the gas become unexcited.  We refer to a disc of such unexcited gas as a ``Frozen Disc''.  Such frozen discs, if they exist, could be entirely invisible in H{\sc i} emission against the CMB, yet absorb light from background objects such as QSOs.

Conversely, gas slabs could be ionised by the intergalactic radiation field unless, in Maloney's (1993) notation:
\begin{equation}
\frac{N_{HI}}{\Delta V} \geq \sqrt{\frac{\phi}{\alpha(T) G \mu}}
\end{equation}
\noindent where $\phi$ describes the ionising field in units of photons\,cm$^{-2}$\,s$^{-1}$, $\alpha(T)$ is the re-combination coefficient and $\mu$ is the total mass density in the disc -- which sets the scale-height of the gas.  For $\alpha \simeq 10^4$ (proximity effect, e.g. Scott et al. 2002), $\mu = 10^{-1}$~g\,cm$^{-2}$ (Milky Way) and $\Delta V\simeq 10$~km\,s$^{-1}$, the column-density limit for stability against ionisation is $N_{HI}\simeq 10^{18.5}$~cm$^{-2}$, an order of magnitude below our observations.

\section{A possible synthesis}

\begin{table}
\caption{Distribution of $T_{HI}$ (in K) for HIDEEP sources}
\begin{tabular}{lrrrrrrrrrrrr}
$T_{HI}$&1-5&5-10&10-15&15-20&20-25&25-30&30-35&35-40&40-45&45-50&50-55&55+\\
Num.&2&12&21&10&11&7&4&4&2&4&5&14\\
\end{tabular}
\end{table}

Note that all the above phenomena depend on $N_{HI}$ and $\Delta V$ in the combination $N_{HI}/\Delta V$.  Furthermore, if we combine Equations (3) and (4) in the optically thin limit we get
\begin{equation}
T_B(v)\simeq T_{bg} + 5.5 \times 10^{-18} \frac{N_{HI}({\rm cm}^{-2})}{\Delta V({\rm km\,s}^{-1})}
\end{equation}
\noindent which suggests we introduce a new quantity
\begin{equation}
T_{HI}(v) = 5.5\times 10^{-18} \frac{N_{HI}({\rm cm}^{-2})}{\Delta V({\rm km\,s}^{-1})}
\end{equation}
\noindent where $T_{HI}(v)$ (``Thigh'') is the surface-brightness, or brightness temperature, of the neutral hydrogen at velocity $v$.

Having introduced $T_{HI}$ we can discuss many effects at once:
\begin{enumerate}
\item{} To be optically thick at 21-cm
\begin{equation}
T_{HI} \geq T_{spin} \hbox{(60 - 100 K)}
\end{equation}
\item{} For global star formation (i.e. gas-disc instability) (Toomre 1964)
\begin{equation}
T_{HI} \geq 5500{\rm K} \frac{\omega(\rm cgs)}{10^{-15}}
\end{equation}
\noindent where $\omega$ = local gas vorticity.
\item{} For discs to be frozen  (Watson \& Deguchi 1984)
\begin{equation}
T_{HI} \leq 7{\rm K}
\end{equation}
\item{}For discs to be ionised (Maloney 1993)
\begin{equation}
T_{HI} \leq 2{\rm K} \sqrt{\frac{\phi/10^4}{\mu/10^{-1}}({\rm cgs})}
\end{equation}
\end{enumerate}

The actual distribution of $T_{HI}$s for the 96 sources in the radio/optical overlap area of HIDEEP is given in Table 1

\begin{figure}
 \centering
 \leavevmode
 \includegraphics[width={0.85\columnwidth}]{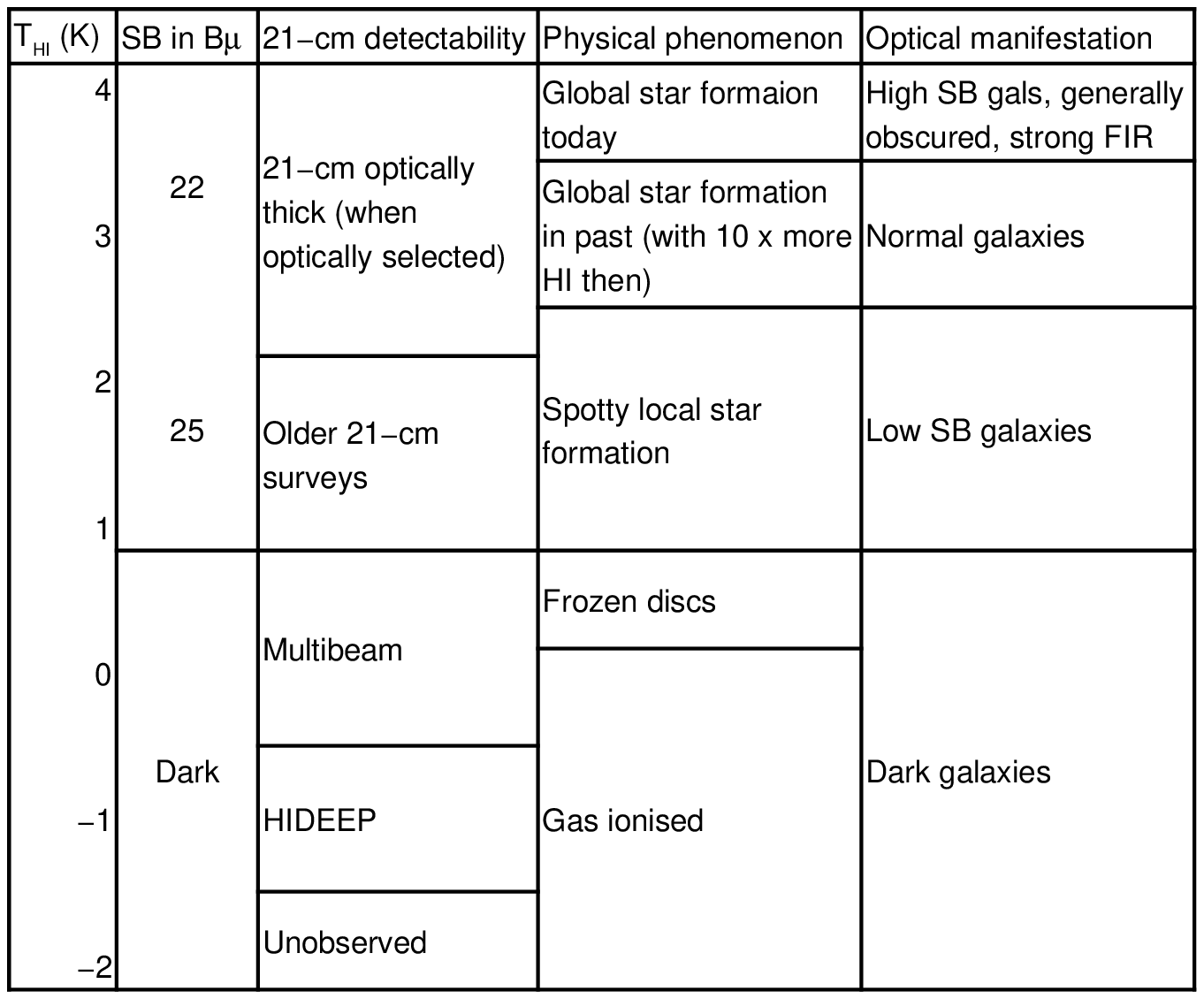}
\caption{Sketch of the 21-cm Universe in terms of $T_{HI}$}
\end{figure}

We note:
\begin{enumerate}
\item{} There is a significant lack of sources below 6~K.  That this cuts in at 7~K, rather than 2~K, suggests that some discs may be frozen while others are ionised.
\item{} The deficit of sources above 60~K suggests that optical depth effects may be important.  We go into this elsewhere (Lang et al. 2002).
\end{enumerate}

Figure 2 is a suggestive sketch of the 21-cm Universe in terms of $T_{HI}$.  At the high $T_{HI}$ end unstable discs underwent global star-formation, leaving behind high SB galaxies today with much lower $T_{HI}$s.  There may be some objects still undergoing global star-formation -- they would be observed as high SB objects with strong FIR outputs.  At the low $T_{HI}$ end would appear frozen discs and discs with their gas ionised -- appearing perhaps as dwarf ellipticals in the optical.

At column-densities above $10^{21.5}$~cm$^{-2}$ molecular hydrogen may become self-shielding to UV radiation and most of the hydrogen will then be in this form.  The H{\sc i} observer would then see only a photo-dissociation region around molecular clouds with an H{\sc i} column-density of $\simeq 10^{20 - 21}$~cm$^{-2}$ (Allen 2001).  If the filling factor for such clouds was close to unity out to well beyond the optical diameter, then the inferred column-density, $N_{HI}^\circ$ would be close to what we observe.  Of course large masses of hidden $H_2$ are implied, apparently sufficient to explain rotation curves.  This fascinating and dramatic idea has been explored by Allen (2001) and cannot apparently be dismissed on observational grounds (see Coombes and Pineau des For\^{e}ts (2000) for a recent comprehensive review).



\begin{chapthebibliography}{1}
\bibitem{author}

Allen, R. J. 2001, ``Gas and Galaxy Evolution'', ed. Hibbard et al. , ASP Conf.\ Ser.\ 240, p331

Coombes, F. \& Pineau des For\^{e}ts 2000, ``Molecular Hydrogen in Space'', Publ.\ Cambridge University Press

Disney, M. J. \& Banks G. D. 1997, PASA, 14, 69

Field, G. 1959, ApJ, 129, 536

Giovanelli, R. \& Haynes, M. P. 1984, AJ, 89, 758

Kulkarni, S. R. \& Heiles, C. 1988, ``Galactic and Extra-Galactic Radio Astronomy'' ed. Kellermann \& Verschuur, second edition, Publ.\ Springer, Berlin, p95

Lang, R. H. et al. 2002, MNRAS, submitted

Maloney, P. 1993, ApJ, 414, 44

Minchin, R. F. 2001, PhD Thesis, Cardiff University

Minchin, R. F. et al. 2002, MNRAS, in preparation

Salpeter, E. E. \& Hoffman G. L. 1996, ApJ, 465, 595

Scott, J., Bechtold, J., Morita, M., Dobrzycki, A. \& Kulkarni, V. P. 2002, ApJ, 571, 665

Toomre, A. 1964, ApJ, 139, 1217

Watson \& Deguchi 1984, ApJ, 281L, 5

\end{chapthebibliography}
\end{document}